\documentclass{article}
\usepackage{graphicx}  
\newcommand{\qx }{$q(x)$}
\newcommand{\Deqx }{$\Delta q(x)$}
\newcommand{\Deqtx }{$\Delta_T q(x)$}

\begin{document}

\title{NEW COMPASS RESULTS ON \\
COLLINS AND SIVERS ASYMMETRIES}
\author{F. Bradamante\\
(on behalf of the COMPASS Collaboration)
\\
\vspace*{0.4cm}
Trieste University and INFN, Via A. Valerio 2,
Trieste, 
Italy}
\maketitle

\begin{abstract}
The study of transverse spin and transverse momentum effects is 
an important part of
the scientific program of COMPASS, a fixed target experiment at the CERN
SPS. For these studies a 160 GeV/c momentum muon beam is scattered on
a transversely polarized nucleon target, and the scattered muon and 
the forward going hadrons
produced in DIS processes are reconstructed and identified in a magnetic
spectrometer. The measurements have been performed on a deuteron target
in 2002, 2003 and 2004, and on a proton target in 2007 and 2010. 
The results obtained for the Collins and Sivers asymmetries from the
data collected in 2010 are here presented for the first time.
They nicely confirm the findings of the 2007 run and allow for reduction of 
the errors by more than a factor of two.
\end{abstract}

\section{Introduction}	

\begin{figure}
\begin{center}
 \includegraphics[width=0.9\textwidth]{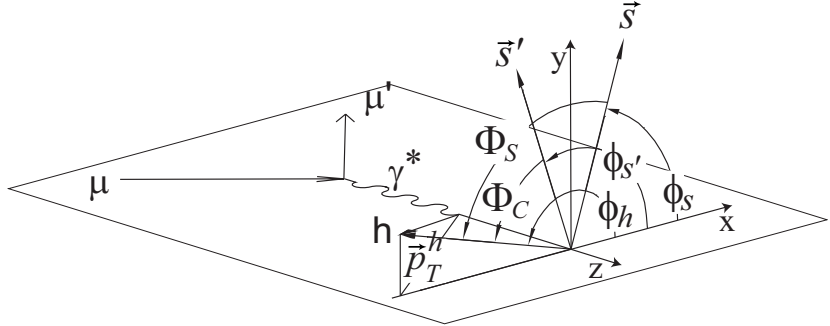}
 \caption{Definition of the Collins and Sivers angles. The vectors 
$\vec{p}_T^{\, h}$ , $\vec{s}$ and $\vec{s'}$ are the hadron
transverse momentum and the spin of the initial and 
struck quarks respectively.
\label{fig:angles}}
\end{center}
  \end{figure}

Transverse spin phenomena in hard processes have been discovered 
and investigated theoretically since 40 years but the field
was vigorously revisited in the 90's,
when a general scheme~\cite{Jaffe:1991kp} of all
leading twist and higher twist parton distribution functions (PDFs)
was worked out.
To fully specify 
the quark structure of the nucleon at the twist-two
level three parton distributions
are necessary, the momentum distributions \qx,
the helicity distributions \Deqx\ and the transverse spin distributions \Deqtx
(or $h_1^q(x)$), where $x$ is the Bjorken variable. 
The latter distribution, known 
as transversity, is chiral-odd and thus not directly observable in 
deep inelastic lepton-nucleon scattering (DIS).
In 1993 Collins suggested that transversity could be measured in semi-inclusive
DIS (SIDIS) thanks to a mechanism involving  in the hadronisation
another chiral-odd function~\cite{Collins:1992kk}, known as the 
``Collins function'' $\Delta_T^0 D_q^h$, i.e. a possible spin dependent
part of the usual fragmentation function $D_q^h$. 
The mechanism is expected to lead to an azimuthal 
transverse spin asymmetry $A_{Coll}$ 
(the ``Collins asymmetry'') in the distribution of the
inclusively produced hadrons.
At leading order this asymmetry can be written as 
\begin{eqnarray}
A_{Coll} = \frac {\sum_q e_q^2 \cdot \Delta_T q \cdot \Delta_T^0 D_q^h}
{\sum_q e_q^2 \cdot q \cdot D_q^h} \, ,
\label{eq:collass}
\end{eqnarray}
and should show up as the amplitude of a  $\sin \Phi_{C}$ 
modulation in the hadron azimuthal distribution.
The Collins angle
$\Phi_{C}=\phi_h+\phi_s-\pi$ is the sum of the azimuthal angles of the 
hadron transverse momentum $\vec{p}_T^{\, h}$ ($\phi_h$)
and of the spin direction of the target nucleon ($\phi_s$)
with respect to the lepton scattering
plane, as measured in the Gamma-Nucleon System. 
Figure~\ref{fig:angles}
 illustrates the choice of the reference system and of the relevant angles.
A non-zero 
Collins asymmetry for the proton was first observed by 
HERMES~\cite{Airapetian:2004tw} implying that the  
Collins fragmentation and the transversity functions are both non-vanishing. 
Independent evidence 
of a non-zero and sizable Collins function came soon after from the 
measurements of a correlation between the azimuthal angles of the hadrons 
in the two jets resulting from the $e^+e^- \rightarrow hadrons$
annihilations into hadrons at high energy as measured by the Belle 
Collaboration ~\cite{Seidl:2008xc}.

If the quarks are assumed to be collinear with the parent nucleon 
(no intrinsic quark
transverse momentum $\vec k_T$), or after integration over $\vec k_T$,
the three distributions \qx,
 \Deqx\ and \Deqtx\ exhaust the information
on the internal dynamics of the nucleon.
However, admitting a finite $\vec k_T$, a total of eight 
transverse momentum dependent (TMD) 
distribution functions are needed for
a full description of the nucleon~\cite{Bacchetta:2006tn}. 
All these functions lead to azimuthal 
asymmetries in the  SIDIS cross-section  and 
can be disentangled measuring their different angular modulations.

In between these TMD PDFs
of particular interest is the Sivers function $\Delta_0^T q$ (or $f_1^q$)
which arises from 
a correlation between the transverse momentum of an unpolarised quark in a 
transversely polarized nucleon and the nucleon polarization 
vector~\cite{Sivers:1989cc} .
Neglecting the hadron transverse momentum with respect to the
fragmenting quark, this $\vec{k}_T$ dependence could cause the ``Sivers
asymmetry'' 
\begin{eqnarray}
A_{Siv} = \frac {\sum_q e_q^2 \cdot \Delta_0^T q \cdot D^h_q}
{\sum_q e_q^2 \cdot q \cdot D_q^h} 
\label{eq:sivass}
\end{eqnarray}
in the angular distribution of the hadrons resulting from the quark
fragmentation.
The Sivers asymmetry is the amplitude of a possible 
$\sin \Phi_{ S}$ modulation in the number of produced hadrons,
where $\Phi_{ S}=\phi_h-\phi_s$ and $\phi_h$ and $\phi_s$
are the same azimuthal angles which enter in definition of $\Phi_{ C}$.
In SIDIS off a transversely polarized target the 
Collins and the Sivers effects can to be disentangled, as 
shown by the COMPASS and the HERMES experiments.

The Sivers function is of particular interest because 
it is odd under time reversal (T).
Due to its T-odd nature and to the T-invariance 
property of the strong interaction soon after being proposed
it was demonstrated that it had to be zero~\cite{Collins:1992kk}.
Ten years later, however, 
in an explicit calculation~\cite{Brodsky:2002cx} it was proved that final state 
interactions in SIDIS arising from gluon-exchange between the
struck quark and the nucleon remnant, or initial state interactions 
in Drell-Yan
processes, can produce a non-zero Sivers asymmetry.
Soon after it was understood~\cite{Collins:2002kn} that taking correctly 
into account
the gauge links in the TMD distributions, time reversal invariance 
does not imply a vanishing Sivers function but rather a sign difference
between the Sivers function measured in SIDIS and the same distribution
measured in Drell-Yan.
Clearly the test of this pseudo-universality  of the T-odd functions
requires their measurement in Drell-Yan process and, first of all, a well
established non-zero signal in SIDIS.

Using a 160 GeV $\mu^+$ beam COMPASS has measured SIDIS on a transversely 
polarized $^6$LiD target in 2002, 2003 and 2004.
In those data no appreciable asymmetries were observed within the accuracy 
of the 
measurements~\cite{Alexakhin:2005iw,Ageev:2006da,Alekseev:2008dn}, a fact 
which is understood in terms of a cancellation between the u- and d-quark 
contributions. 
The COMPASS data are still today the only SIDIS data
ever taken on a transversely polarized deuteron target, and provide 
important constraints to the contribution of the d-quark.
Together with the HERMES data on a transversely polarized proton target
and the $e^+e^- \rightarrow hadrons$ Belle data they  
allowed for the first
global analysis and the first extraction of the transversity distributions 
and of the 
Sivers functions for the u- 
and d-quarks~\cite{Efremov:2008vf,Anselmino:2008jk,Anselmino:2008sga}. 
In 2007 COMPASS measured for the first time SIDIS on a transversely polarized 
proton (NH$_3$) target
\footnote{For a comprehensive review of recent experiments and theoretical 
developments see f.i.~\cite{Barone:2010zz}.}. 
The results~\cite{Alekseev:2010rw} for the Collins asymmetry were in 
nice agreement with those of 
HERMES~\cite{Airapetian:2009ti,Airapetian:2010ds}, 
while the Sivers asymmetry turned out 
to be somewhat smaller. 
Understanding the reasons for this difference was a strong motivation 
for a new proton run, and the entire 2010 data taking period, 
from June to November, 
was dedicated to such a measurement. 
We have just recently finished the 
data analysis and I have the great pleasure to present here for the first 
time the new results.

\section{The COMPASS Experiment}	

The COMPASS spectrometer~\cite{Abbon:2007pq} is in operation in the 
North Hall (Hall 888) of CERN since 2002.
To ensure large angular acceptance and dynamical range
it is a two-stage magnetic spectrometer, and to cope with the
different requirements of location accuracy and rate capability at different 
angles 
it uses a variety of tracking detectors.
Particle identification is provided by a large acceptance RICH detector, 
by calorimeters,
and by muon filters.

In 2010 the spectrometer configuration was very similar 
to the one which was used in 
2007. The main difference is the addition of a new triggering system for 
large-angle muons, based on two large area scintillator counter hodoscopes 
with 32 horizontal bars each and a suitable coincidence matrix to provide 
target pointing in the non-bending vertical plane. 

The polarized target consists of three cylindrical cells, 4 cm diameter:
one central cell,  60 cm long, and two outer ones, each 30 cm long, separated 
by 5 cm.
Neighboring cells are polarized in opposite directions, so that data from both
spin directions are recorded at the same time.

To polarize the target material and to hold the polarization in the 
longitudinal 
mode the target is located along the axis of a solenoid which provides a field 
of 2.5 T with an excellent homogeneity
of $\pm2 \cdot 10^{-5}$ over the whole target volume.
The magnet is also equipped with a
set of saddle-shaped coils which can provide a 0.6 T vertical field,
which is used either to rotate the target nucleon spin or to hold
the polarization vertical for the transversity measurements. 
The target material 
(NH$_3$) is first longitudinally polarized in the
solenoidal field with the method of dynamical nuclear
polarization (DNP).
The maximum polarization value achievable in the three cells
ranges between 0.80 and 0.90, depending slightly on the sign of the
polarization and on the cell location.
About 48 hours are necessary to reach 95\% of the maximal polarization.
When the desired polarization value is reached, the radio frequency system 
is switched off, the target
spins get frozen, the magnetic field is lowered to 0.6 T and adiabatically 
rotated 
to the vertical direction by suitably varing the solenoid
and the dipole fields.
In the frozen spin mode and with the holding field at its operational value 
the relaxation time of the polarization exceeds 3000 hours.

In 2010 data have been taken at a  mean beam intensity of 
$2\cdot 10^8 \; \mu$/spill,
for a spill length of $\sim 10$ s every 40 s.
About $37\cdot 10^9$ events, corresponding to 1.3 PB of data, have been 
collected,
in twelve separate periods.
In each period, after 4-5 days of data taking, a polarization reversal 
was performed 
by changing the microwave frequencies in the three cells, always keeping 
the central cell
oppositely polarized with respect to the external ones.

\begin{figure}
\begin{center}
 \includegraphics[width=0.48\textwidth]{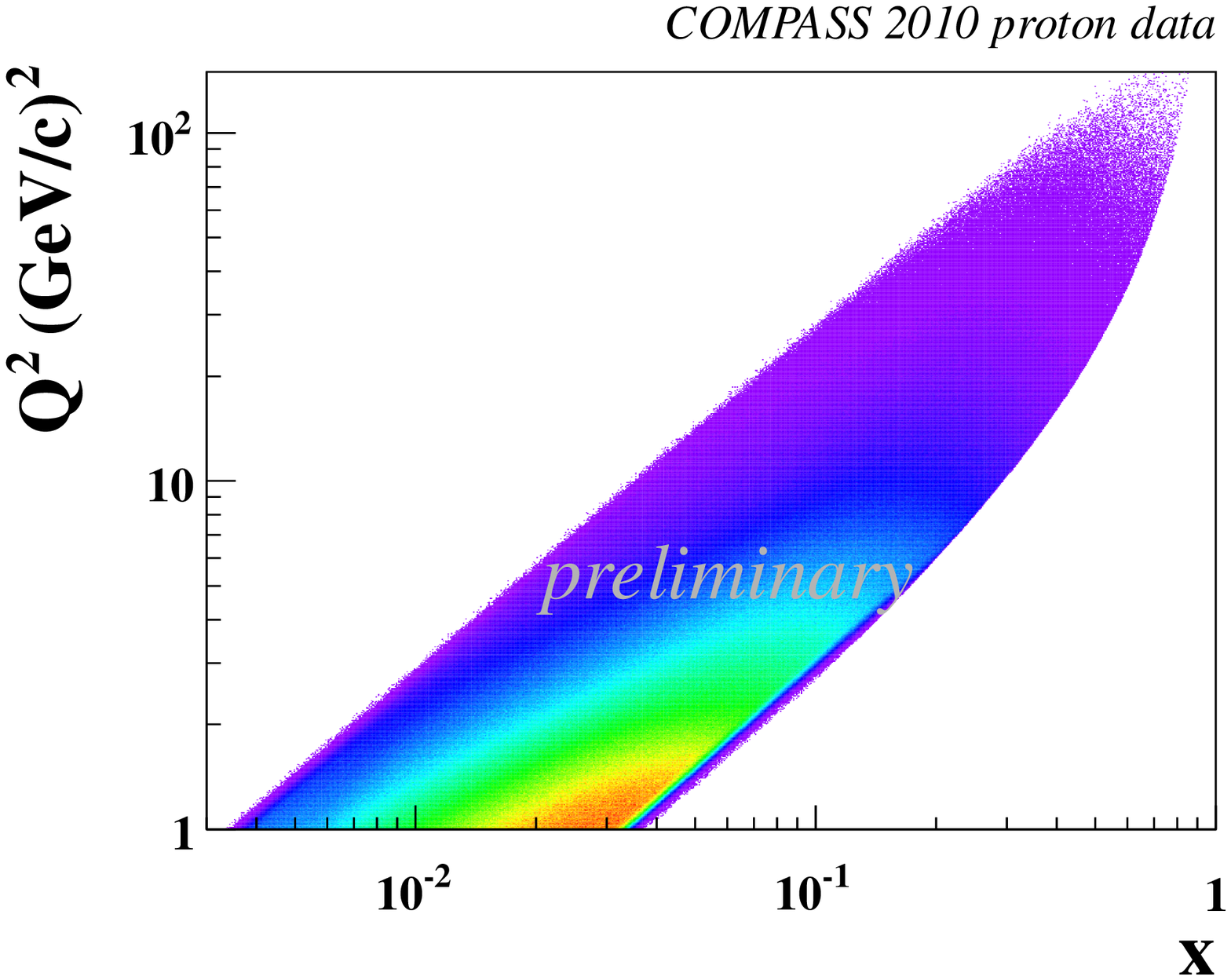}
\hspace*{0.2cm}
 \includegraphics[width=0.48\textwidth]{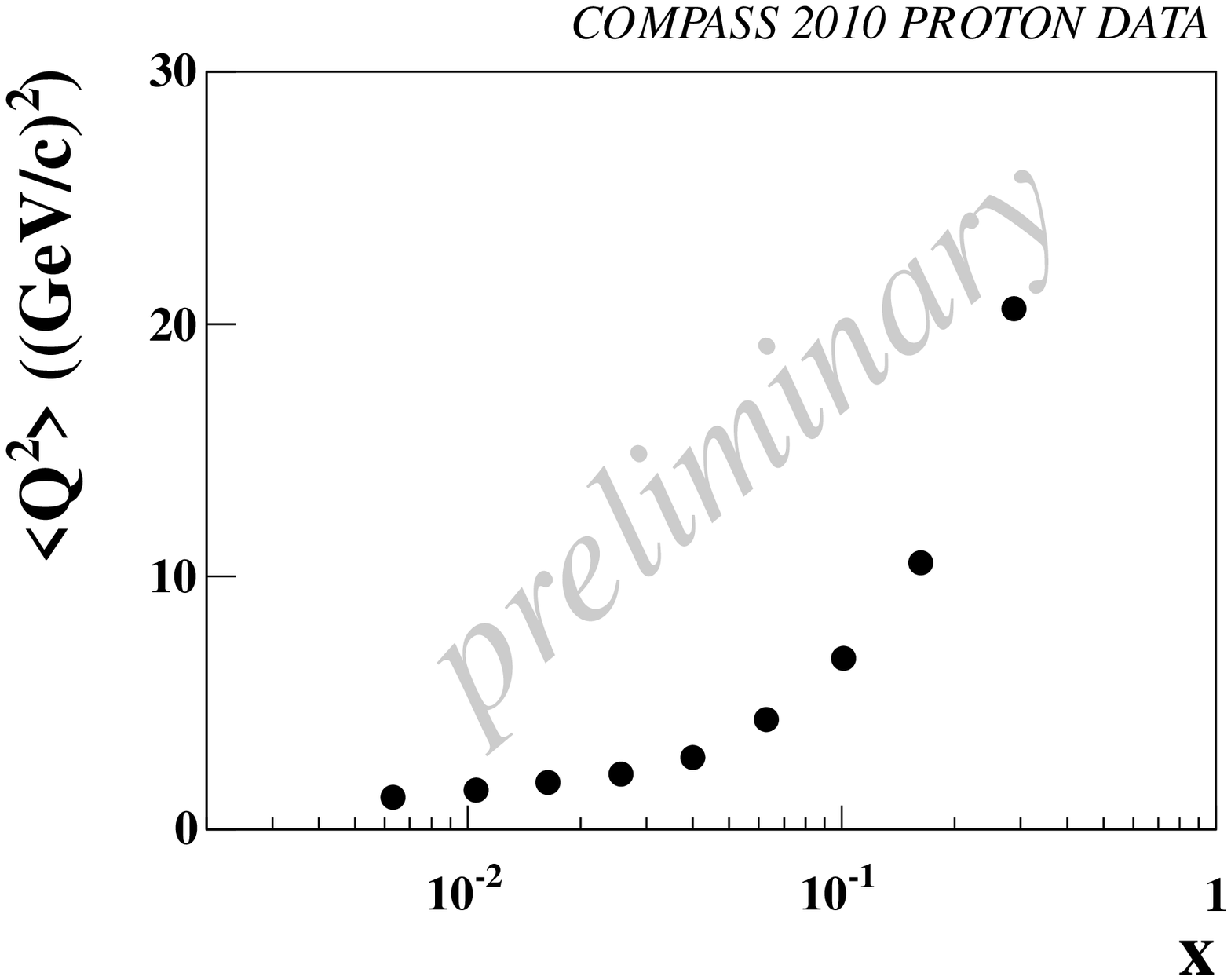}
\hfill
\vspace*{-0.5cm}
 \caption{Left: distribution of the DIS events in 
the $x-Q^2$ plane.
Right: mean $Q^2$ values in the different $x$-bins.
\label{fig:kin}}
\end{center}
  \end{figure}

In the data analysis, 
in order to ensure a DIS regime, only events with a photon virtuality $Q^2>1$
(GeV/c)$^2$, 
a fractional energy of the virtual photon $0.1<y<0.9$, and 
a mass of the hadronic final  state system $W>5$ GeV/c$^2$ are considered. 
The charged hadrons are required to have at least 0.1 GeV/c transverse
momentum $p_T^h$ with respect to the virtual photon direction and a 
fraction of the available  energy $z>0.2$.
Within these cuts, we are left with about $16\cdot 10^7$ DIS events, 
and $8\cdot 10^7$ hadrons.
Figure~\ref{fig:kin} shows at the left the distribution of the DIS events in 
the $x-Q^2$ plane, while the mean $Q^2$ values in the $x$-bins in which we
give the results for the asymmetries are shown in the plot at the right.

The transverse spin asymmetries are obtained by comparing, cell by cell,
the azimuthal distributions of the detected hadrons as measured in the first
half of a period with the corresponding distributions of the second
half.
Since the two sets of data are taken typically one week apart, the stability of
the apparatus is crucial and has been carefully checked.
All the tests on the data quality and stability have not revealed any 
systematic effects, all the twelve periods give compatible
results, and all of them have  been used for 
the extraction of the final asymmetries.

\section{Results}

The Collins and Sivers asymmetries have been evaluated for positive and 
negative hadrons 
in bins of the three kinematic variables $x$, $z$ and $p^h_T$.
The raw asymmetries have been extracted for each  data taking period
and have been divided
by the dilution factor, the target polarization and, in the case of 
the Collins analysis,
by the $D_{NN}$ kinematical factor.
The dilution factor of the ammonia target has been evaluated in each 
$x$ bin; it increases with $x$ from 0.14 to 0.17, and it has been
assumed constant and equal to 0.15 in the $z$ 
and $p^h_T$ bins.
The target polarization was measured individually for each cell and each 
period, with a 5\% uncertainty.
Finally, the weighted mean of the asymmetries measured in each period has
been performed.
The  systematic errors have been evaluated as a fraction of the  
statistical error
and are 0.5 for both the Collins and Sivers asymmetries.

\begin{figure}
\begin{center}
 \includegraphics[width=0.95\textwidth]{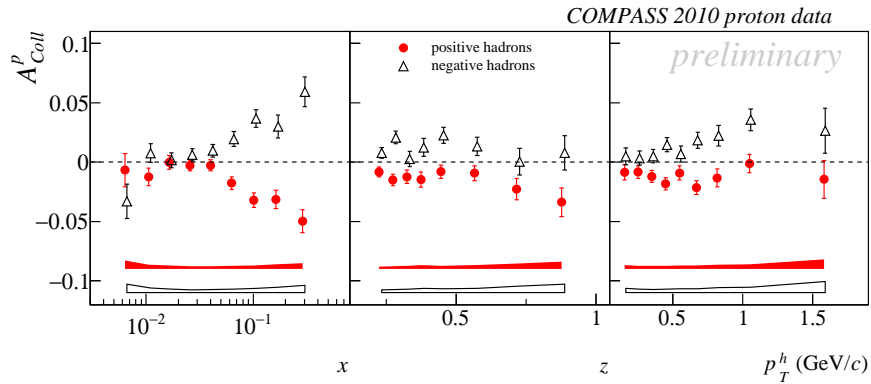}
\vspace*{-0.4cm}
 \caption{Collins  asymmetry as function of 
$x$, $z$, and $p_T^h$, for positive and negative hadrons. 
\label{fig:res_col}}
\end{center}
  \end{figure}

\begin{figure}
\begin{center}
 \includegraphics[width=0.95\textwidth]{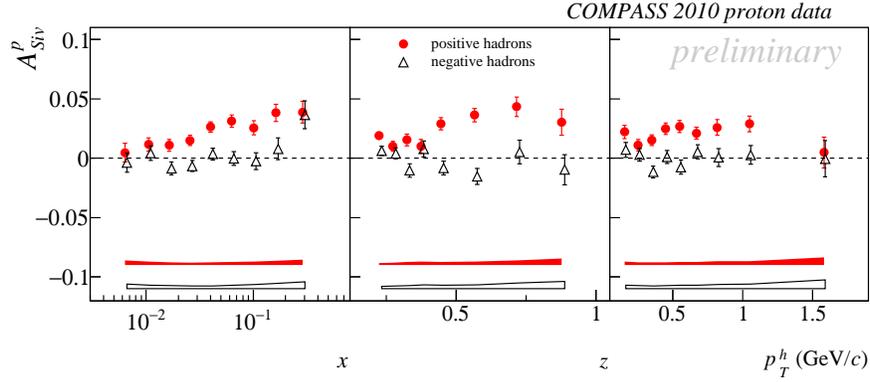}
\vspace*{-0.4cm}
 \caption{Sivers asymmetry as function of 
$x$, $z$, and $p_T^h$, for positive and negative hadrons. 
\label{fig:res_siv}}
\end{center}
  \end{figure}

The asymmetries are given as function of 
$x$, $z$, and $p_T^h$, for positive and negative hadrons,
shown in fig.~\ref{fig:res_col} and~\ref{fig:res_siv}.
The error bars are only statistical, while the bands indicate the size of 
the systematic errors.
As is clear from fig.~\ref{fig:res_col} the Collins asymmetry 
has a strong $x$ dependence. It is  compatible with zero
at small $x$ and increases up to 0.10 in the valence region ($x>0.1$).
The values agree both in magnitude and in sign
with our previous measurements~\cite{Alekseev:2010rw}, as well as 
with the measurements of 
HERMES~\cite{Airapetian:2010ds},
which have been performed at a considerably lower electron beam energy.

Figure~\ref{fig:res_siv} shows our new results for the Sivers asymmetries. 
Again, there is an excellent agreement with our published results from 
the 2007 run, with a considerable reduction of the error bars (more than 
a factor of two). 
The asymmetry is definitely positive for positive 
hadrons and compatible with zero for negative hadrons. At variance with 
the Collins asymmetry, the Sivers asymmetry stays positive even for 
very small $x$-values, in the region of the sea. 
Moreover, very much as 
was the case for the 2007 data, the measured asymmetries are definitely 
smaller than the corresponding ones measured by HERMES. 
To understand the reason we have enlarged the kinematic domain, namely 
we have looked at the events with smaller $y$ values (in the
interval $0.05<y<0.1$) and at the hadrons with 
smaller $z$ values ($0.1<z<0.2$).

\begin{figure}
\begin{center}
 \includegraphics[width=0.95\textwidth]{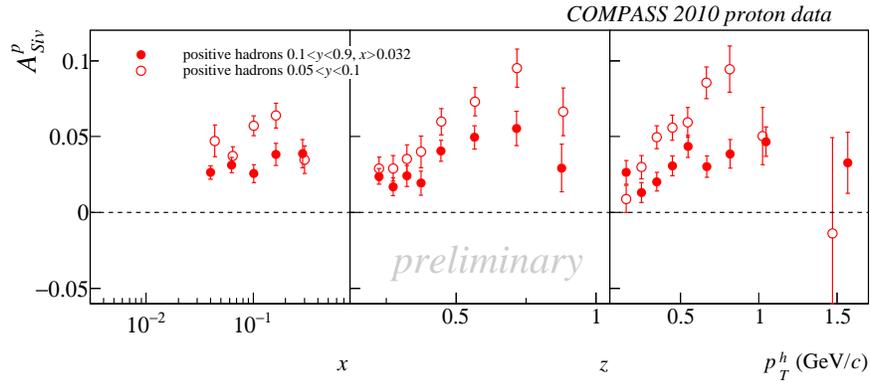}
\vspace*{-0.4cm}
 \caption{Sivers asymmetry for positive hadrons as function of 
$x$, $z$, and $p_T^h$, for $x>0.032$.
The full points refer to the $0.1<y<0.9$ sample, the open points 
to the $0.05<y<0.1$ sample.
\label{fig:res_siv_ly}}
\end{center}
  \end{figure}

Figure~\ref{fig:res_siv_ly} shows the Sivers asymmetry for positive 
hadrons as function of $x$, $z$, and $p_T^h$ for small $y$ 
values ($0.05<y<0.1$) 
as compared with the ``standard'' sample ($0.1<y<0.9$). 
Since at small $y$ there are no low-$x$ data, only data with $x>0.032$
are plotted. A clear increase of the Sivers asymmetry is visible for the low-y
data.
This strong effect  could be associated with the smaller values of $Q^2$
and/or with the smaller values of $W$, the invariant 
mass of the hadronic system.
The ``standard sample'' ($y>0.1$) corresponds 
to $W > 5$ GeV,
while in the range $0.05<y<0.1$ 
the $W$ values are as low as $\sim$3 GeV. 
While a $Q^2$ dependence is expected and has been calculated~\cite{aybat:2011},
no dependence on $y$ (nor
on $W$) is foreseen. 
A similar correlation, statistically less significant, was
already noticed in our published 2007 proton data. 
Clearly this point needs further investigation.
No particular trend is observed for the case of the Sivers asymmetry of the 
negative hadrons, which are compatible with zero for the standard sample 
and stay 
compatible with zero at small $y$.

\begin{figure}
\begin{center}
 \includegraphics[width=0.95\textwidth]{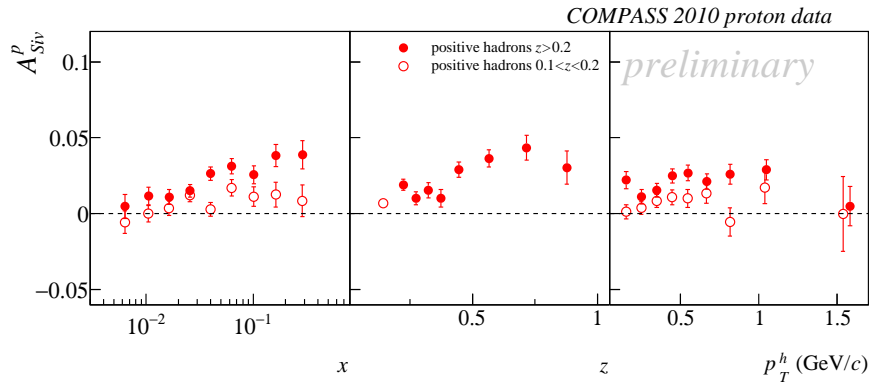}
\vspace*{-0.4cm}
 \caption{Sivers asymmetry for positive hadrons as function of 
$x$, $z$, and $p_T^h$.
The full points refer to the  $z>0.2$ sample, the open points 
to the $0.1<z<0.2$ sample.
\label{fig:res_siv_lz}}
\end{center}
  \end{figure}

We have also investigated the trend of  the Collins and Sivers 
asymmetries at low $z$.
Our standard hadron selection requires $z>0.2$, to stay well 
separated from the target hadronization. 
In the region $0.1<z<0.2$ the Collins asymmetries 
decrease only slightly in size,
on the other hand the effect is again sizable for the Sivers 
asymmetry for the positive hadrons, and not visible for negative
hadrons. 
Figure~\ref{fig:res_siv_lz} 
compares the data in the ``standard sample'' ($z>0.2$) with the events 
in the region ($0.1<z<0.2$) and the decrease of the Sivers asymmetry 
for positive hadrons is impressive.
Also this effect looks quite interesting and will be further
investigated.

\section{Conclusions}

Transverse spin and transverse momentum effects in lepton nucleon scattering 
offer new tools to unveil the structure of the nucleon.
COMPASS contributed its share by executing a first round of exploratory 
measurements scattering 160 GeV/c muons on transversely polarized
deuteron and proton targets. 
The proton data are particularly important because of the relatively large 
Collins and Sivers asymmetries which HERMES and COMPASS have observed 
for the first time.
The COMPASS measurements are quite useful in assessing the leading twist 
nature of the effects since the asymmetries are non-zero mostly in the 
valence region, where the $Q^2$ values of the measurements are a 
factor of 3 to 4 larger than those of HERMES.
By now, the amount of data which has been produced by HERMES, COMPASS and 
Belle is already  impressive, and time
has come for new phenomenological global analyses.
Also, many more results are in the COMPASS pipe-line and should
appear soon.
The analysis of the Collins and Sivers asymmetries I have shown
is essentially over, but work is still ongoing on the corresponding
asymmetries for identified hadrons and on the kinematical dependence of
the effect.
Work is ongoing also to extract the other six transverse spin 
dependent asymmetries
which are present in the expression of the SIDIS cross-section.
Also in this case it is of interest to consider separately
pions and kaons.

In the long term, the investigation of the spin structure of the
nucleon in SIDIS 
will necessitate a major investment, to build a high luminosity 
electron-proton collider in which polarized electrons and polarized proton 
will collide at high energy. Projects are ongoing since some time at JLAB, at 
BNL, and, more recently, in Europe, where ideas to use the HESR antiproton 
storage ring of FAIR at GSI to store polarized protons are
being put forward. 
The construction of a new polarized electron ring of 
suitable energy is not for free, but feasibility studies are ongoing. 
In the meantime, only JLAB and COMPASS can contribute
to this field. JLAB experiments are unbeatable in statistical precision,
but the interpretation of the data requires also measurements at high $Q^2$
which can only be performed using high energy beams. From this point of
view, I think COMPASS should stay on the stage for several years, beyond
the presently approved Drell-Yan~\cite{denisov:2011}
and DVCS~\cite{dhose:2011} measurements, profit
of the accelerator complex upgrade which is being carried on at CERN and
thus increase its luminosity, and bridge the colleagues who are interested
in this field across the time gap from now to the day the future collider
will enter into operation.

\end{document}